\documentclass[12pt,a4paper]{article}
\usepackage{graphics,epsfig,rotating,amssymb}
\begin{document}
\newcommand{\goo}{\,\raisebox{-.5ex}{$\stackrel{>}{\scriptstyle\sim}$}\,}
\newcommand{\loo}{\,\raisebox{-.5ex}{$\stackrel{<}{\scriptstyle\sim}$}\,}

\begin{center}
{\large \bf Multifragmentation reactions and
properties of hot stellar matter at sub-nuclear densities.}
\end{center}
\vspace{0.5cm}
\begin{center}
{\Large A.S.~Botvina$^{a}$ and I.N.~Mishustin$^{b,c}$}\\
\end{center}
\begin{center}
{\it
 $^a$Institute for Nuclear Research, Russian Academy of Sciences, 117312 Moscow,
 Russia\\
 $^b$Frankfurt Institute for Advanced Studies, J.W. Goethe University, D-60438
     Frankfurt am Main, Germany\\
 $^c$Kurchatov Institute, Russian Research Center, 123182 Moscow, Russia\\
}
\end{center}
\normalsize

\vspace{0.3cm}
\begin{abstract}
We point out similarity of thermodynamic conditions reached in
intermediate-energy nuclear collisions and in supernova explosions.
We show that a statistical approach, which has been previously applied
for nuclear multifragmentation reactions, can be very useful
for description of the electro-neutral stellar matter.
Then properties of hot unstable nuclei extracted from analysis of
multifragmentation data can be used for construction of a realistic
equation of state of supernova matter.
\end{abstract}

\vspace{0.3cm}

{\bf PACS: 25.70.Pq , 26.50.+x , 21.65.+f }

 \vspace{0.5cm}

A type II supernova explosion is one of the most spectacular events in
astrophysics, with huge energy release of about $10^{53}$ erg or several
tens of MeV per nucleon \cite{Bethe}.
When the core of a massive star collapses, it reaches densities several
times larger than the normal nuclear density $\rho_0=0.15$ fm$^{-3}$.
The repulsive nucleon-nucleon interaction gives rise to a bounce-off
and creation of a shock wave propagating through the in-falling stellar
material.
This shock wave is responsible for the ejection of a star envelope that
is observed as a supernova explosion.

During the collapse and subsequent explosion the temperatures
$T\approx (0.5\div 10)$ MeV and baryon densities $\rho_B \approx
(10^{-5}\div 2) \rho_0$ can be reached. As shown by many
theoretical studies, a liquid-gas phase transition is expected in
nuclear matter under these conditions. It is remarkable that
similar conditions can be obtained in energetic nuclear collisions
studied in terrestrial laboratories. This fact gives a ground to
use well established models of nuclear reactions, after certain
modifications, for describing matter states in the course of
supernova explosions. On the other hand, the supernova physics
stimulates investigations of specific reaction mechanisms and
exotic nuclei.

As demonstrated by several authors (see e.g. \cite{Janka,Thielemann}), present
hydrodynamical simulations of the core collapse do not produce successful
explosions, even when neutrino heating and convection effects are included.
On the other hand, it is known that nuclear composition is extremely important
for understanding the physics of supernova explosions. In particular, the
weak reaction rates and energy spectra of emitted neutrinos are very sensitive
to the presence of heavy nuclei (see e.g. \cite{Ring,Hix,Langanke,Horowitz}).
This is also true for the equation of state (EOS) used in hydrodynamical
simulations since the shock strength is mainly diminished due to the
dissociation of heavy nuclei.

The EOS of supernova matter is under investigation for more than 20 years.
A most popular EOS which is presently used in supernova simulations
was derived in refs. \cite{Lamb,Lattimer}.
However, it is based on properties of cold or nearly cold nuclei ($T \loo 1$ MeV)
extracted from nuclear reactions at low (solid state target) density
(i.e., $\rho_B \sim 10^{-14}\rho_0$). On the other hand, properties of nuclei
may change significantly at high temperatures, baryon and lepton densities
associated with supernova explosions.
For more realistic description of supernova physics one should certainly
use experience accumulated in recent years by studying intermediate-energy
nuclear reactions. In particular, multifragmentation
reactions provide valuable information about hot nuclei in dense environment.
We believe that properties of equilibrated transient systems produced
in these reactions are similar to those expected in supernova explosions.
Another shortcoming of calculations presented in refs. \cite{Lamb,Lattimer})
is that they reduce an ensemble of hot nuclei to a single "average nucleus".
This is a crude approximation which may strongly distort the true statistical
ensemble. Therefore, an urgent task now is to derive a more realistic EOS of hot and
dense stellar matter which can be applied for a broad range of thermodynamic
conditions. The first step in this direction was made in our previous paper
\cite{Botvina04}.
A similar model was also used ref. \cite{Japan} where, however,
only cold nuclei in long-lived states were considered.

In the supernova environment, as compared to nuclear reactions,
new important ingredients should be taken into consideration. The
matter at stellar scales is electrically neutral, therefore,
electrons must be included to balance a positive nuclear charge.
Energetic photons are also present in hot matter and they can
change the nuclear composition via photo-nuclear reactions.
Finally, the flavor content of matter can be affected by a strong
neutrino flux from the newly-born protoneutron star. The crucial
question for theoretical description is what degree of
equilibration is reached in different reactions. Our estimates
show that at temperatures and densities of interest the
characteristic reaction times for nuclear interactions vary within
the range from 10 to 10$^6$ fm/c, that is very short compared to
the characteristic time of the explosion (about 100 ms
\cite{Janka}). The assumption of nuclear equilibration is fully
justified for these conditions, and therefore, the nuclear
composition can be determined from a statistical model. The rate
of photo-nuclear reactions depends strongly on the density, and at
very low densities, less than $10^{-5}-10^{-6} \rho_0$, these
reactions are more efficient than nuclear interactions.
The weak
interactions are much slower. It is most likely that at high
densities, $\rho_B\goo 10^{-3} \rho_0$, neutrinos are trapped in a
nascent neutron star, but at lower densities they stream freely
from the star. The weak processes are entirely responsible for the
neutrino and electron content of the matter. For example, the
electron capture may not be in equilibrium with nuclear reactions
at small densities \cite{Langanke03}. Therefore, an adequate
treatment is needed to discriminate various conditions with
respect to the weak reactions \cite{Botvina04}. We conclude that
the statistical approach can be applied for nuclear reactions, but
possible deviations from equilibration for weak and
electromagnetic interactions should be explicitly taken into
account.



Statistical models have been proved to be very successful in nuclear
physics. They are used in situations when an equilibrated source can
be defined in the reaction. The most famous example of such a
source is the 'compound nucleus' introduced by Niels Bohr in 1936.
The standard compound nucleus picture is valid only at low excitation
energies when evaporation of light particles and fission are the dominating
decay channels. However, this concept cannot be applied at high excitation
energies, $E^* \goo$ 2--3 MeV/nucleon, when the nucleus breaks up
fast into many fragments. Several versions of the statistical approach
have been proposed for the description of such multifragmentation reactions
(see e.g. \cite{Randrup,Gross,SMM}).
As was demonstrated in many
experiments (see e.g. \cite{ALADIN,EOS,ISIS,MSU,INDRA,FAZA}), an equilibrated
source can be formed in this case too, and statistical models are generally
very successful in describing its properties. Furthermore, systematic studies
of such highly excited systems have brought important information about
the nuclear liquid-gas phase transition \cite{Pochodzala,Dagostino2}.

As a basis for our study we take the Statistical
Multifragmentation Model (SMM), see a review \cite{SMM}.
Presently, the SMM is the only model of multifragmentation which
can be applied in the thermodynamical limit for infinite systems
\cite{Bugaev}. This makes possible to use it for astrophysical
conditions. The model assumes statistical equilibrium at a
low-density freeze-out stage. It considers all break-up channels
composed of nucleons and excited fragments taking into account the
conservation of baryon number, electric charge and energy. Light
nuclei with mass number A $\leq 4$ are treated as elementary
particles with only translational degrees of freedom ("nuclear
gas"). Nuclei with A $> 4$ are treated as heated liquid drops.
In this way one may study the nuclear liquid-gas coexistence in
the freeze-out volume. The Coulomb interaction of fragments is
described within the Wigner-Seitz approximation. Different channels
$f$ are generated by Monte Carlo sampling according to their
statistical weights, $\propto \exp{S_f}$, where $S_f$ is the
entropy of channel $f$. After the break-up the Coulomb acceleration
and the secondary de-excitation of primary hot fragments is taken
into account.




An important advantage of the SMM is that it includes all break-up
channels ranging from the compound nucleus to
vaporization\footnote{By vaporization we mean all channels which
contain only light particles (A$\leq 4$).}, and one can study the
competition between them. As was shown already in first
publications \cite{SMM,SJNP85}, multifragmentation dominates over
compound nucleus at high excitation energies, and must be taken
into account in order to explain multiple fragment production.
Most clearly this is demonstrated in Fig.~1 where we show
entropies of two extreme disintegration modes of $^{238}{\rm U}$,
namely the compound nucleus (CN) and the vaporization (V)
channels, as functions of excitation energy. One can clearly see
that the CN channel dominates at low excitation energies but the V
channel wins at excitation energies above 12 MeV/nucleon. However,
in fact the CN channel dies out at much lower excitation energies,
2-3 MeV/nucleon, when the channels containing a heavy residue (HR)
and/or several intermediate-mass fragments (IMFs: $4<$A$<50$) have
a higher entropy. The entropy for the whole ensemble of
fragmentation channels is also shown in Fig.~1 (solid line). One
can see, for instance, that at 5 MeV/nucleon this ensemble has a
0.2 higher entropy per nucleon than the CN channel. This means
that the relative probability of the CN channel is $\exp{\Delta
S}= \exp{(-0.2\cdot 238)}\approx 2\cdot 10^{-21}$. This is a
general trend, and it can not be reversed by another description
of the compound nucleus \cite{Sobotka}.
\begin{figure} [tbh]
\vspace{2cm}
\hspace{3cm}
\includegraphics[width=10cm]{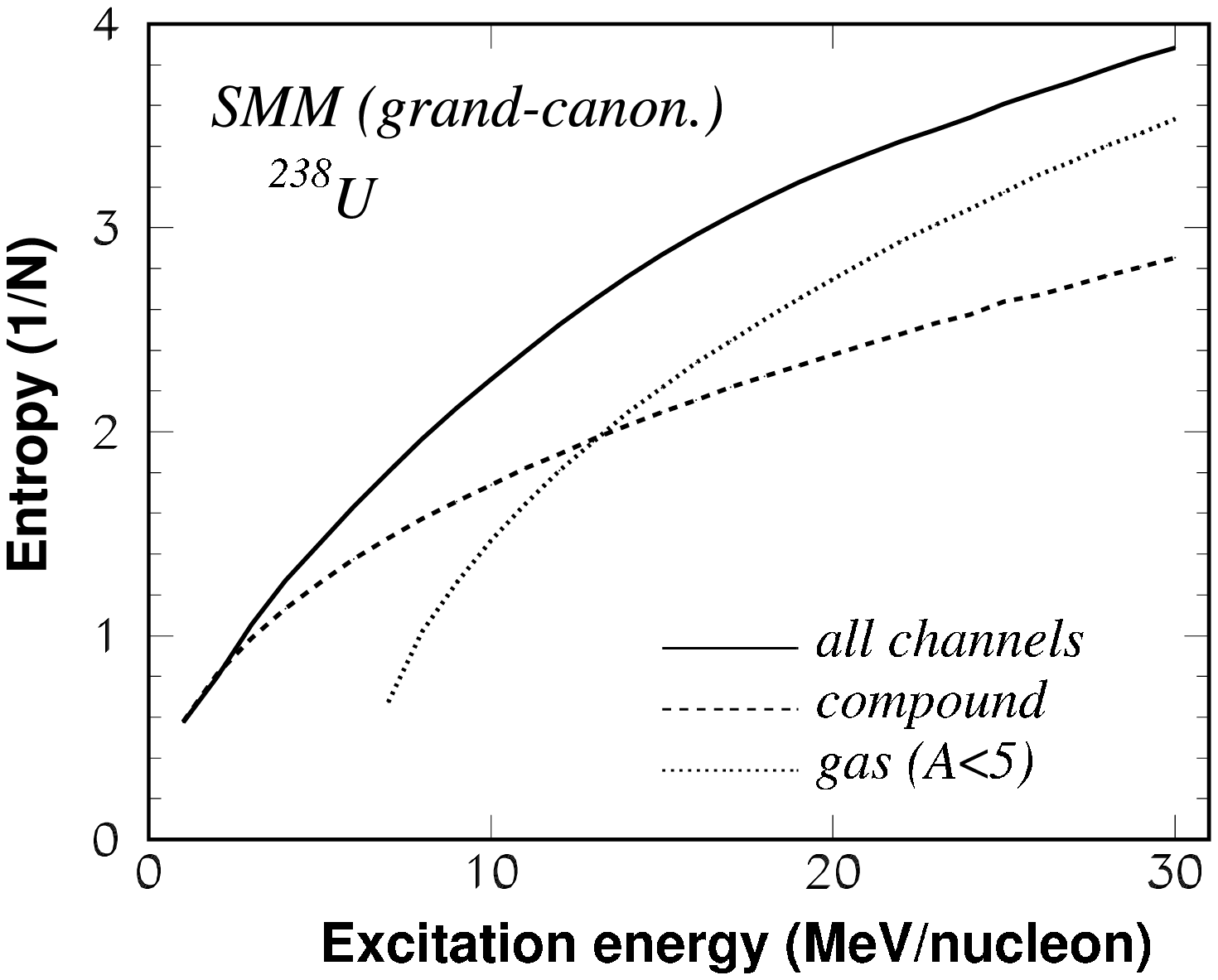}
\caption{\small{ Entropy per nucleon for different disintegration
channels of $^{238}$U as a function of excitation energy per
nucleon. Calculations are performed within the grand-canonical
version of the SMM (details see in refs. \cite{SMM,SJNP85}), at
the freeze-out density $\rho=\rho_0/3$. Dashed and dotted lines
correspond to compound nucleus and vaporization (A$\leq$4)
channels, respectively. The total entropy, obtained by summing
over all break-up channels, is shown by solid line. }}
\end{figure}
We emphasize that the difference between the solid and dotted
lines in Fig.~1 is caused entirely by the presence of hot heavy
and intermediate-mass fragments. From the figure one can conclude
that their contribution remains significant even at very high
excitation energies, up to 30 MeV/nucleon, and nuclear entropies
up to 4 units per baryon. This observation is very relevant for
the physics of supernova explosions since the survival of
relatively heavy nuclei may help reviving the shock wave. Indeed,
if nuclei can exist under such extreme conditions, the energy
losses for dissociation of initial nuclei will be considerably
reduced.

For astrophysical applications it is important that the SMM,
besides fragment partitions, can describe well the neutron content
of fragments \cite{ALADIN}. Generally, in the case of neutron-rich
sources, the SMM predicts neutron-rich hot primary fragments. As
calculations show  \cite{Botvina01,Shetty05}, they keep a part of
their neutron excess even after de-excitation. New experiments
give evidence for production of such unusual neutron-rich nuclei,
which, however, should be  quite common for supernova matter. For
example, Fig.~2 shows the neutron-to-proton ratio (N/Z) for
fragments produced in fragmentation of $^{238}$U with energy 1
GeV/nucleon on Pb and Ti targets \cite{FRS}. The experiment was
performed with Fragment Separator (FRS) at GSI. Fission and
spallation fragments were excluded from the analysis in order to
guarantee selection of multifragmentation-like events. One can see
that the observed neutron content of fragments with $Z<60$ is
larger than expected from the standard EPAX parametrization, which
is a result of spallation-like processes considered previously as
the main mechanism for production of these fragments. Moreover, at
Z$<$30 it becomes larger than the neutron content of stable
nuclei. These results are fully consistent with previous findings
of the ALADIN collaboration \cite{ALADIN}.

\begin{figure} [tbh]
\vspace{-1cm}
\hspace{2cm}
\includegraphics[width=10cm]{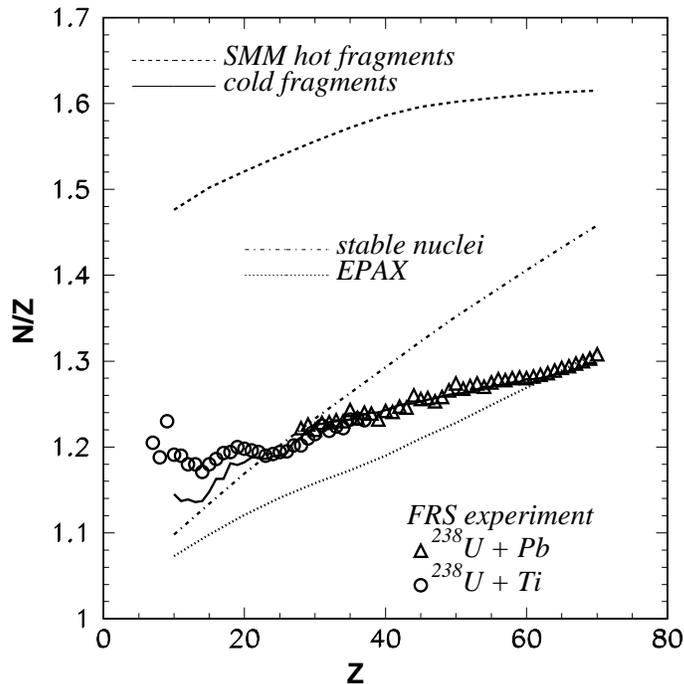}
\caption{\small{
Mean neutron-to-proton ratio versus charge of fragments produced in
multifragmentation-like break-up of $^{238}$U with energy 1 GeV/nucleon
on Pb and Ti targets. Points are experimental data obtained on Fragment
Separator at GSI \cite{FRS}. The SMM calculations are shown by dashed
(primary hot fragments) and solid (fragments after secondary de-excitation)
lines. Dash-dotted line corresponds to stable nuclei, dotted line is the
EPAX phenomenological parametrization for nuclei produced by spallation.
}}
\end{figure}
In Fig.~2 we demonstrate that the SMM can quantitatively describe
this trend by predicting at the same time primary hot fragments
with very large $N/Z$. In recent years the interest to isospin
degrees of freedom has considerably increased and strong
experimental programs to study isospin-asymmetric nuclei exist now
at GSI (Germany), TAMU (USA), NSCL/MSU (USA), INFN (Italy), GANIL
(France) and other laboratories. These studies will give
information about the symmetry energy in hot nuclei, e.g. via the
isoscaling phenomenon \cite{Botvina02}. Presently, there are
indications of essential decrease of the symmetry energy in hot
nuclei \cite{Shetty05,LeFevre}. A small discrepancy shown in
Fig.~2 between the theory and experiment for fragments with Z$<20$
can be attributed to this effect:   lower symmetry energy in the
beginning of the secondary evaporation cascade can increase the
neutron richness of final cold fragments \cite{nihal}. This
possibility should be kept open when adjusting the model
parameters for best description of hot neutron-rich nuclei
produced in nuclear reactions. Then these results can be used for
constructing a reliable equation of state for nuclear matter at
supernova conditions.



According to present knowledge, supernovae of type II are
trigerred by the collapse of massive stars after formation of a
big enough iron core. After the bounce of infalling matter off a
dense core a shock wave is generated, it propagates outwards
leaving behind a highly compressed and heated matter. This matter
consists of various nuclear species as well as free neutrons,
protons, electrons, photons and possibly trapped neutrinos
\cite{Bethe}. Calculation of the equation of state of supernova
matter presents a challenge to modern nuclear physics. Due to the
presence of leptons the EOS is essentially modified as compared
with the case of pure nuclear matter. At densities below
0.5$\rho_0$ uniform nuclear matter breaks into fragments and
nucleons immersed in the uniform electron-positron plasma. The EOS
should cover a broad range of baryon densities and temperatures
expected during the core collapse and subsequent explosion.

In order to describe nuclear composition of supernova matter we
have modified the SMM by including electrons and neutrinos in the
statistical ensemble \cite{Botvina04}. The model was formulated
for different assumptions concerning the weak reaction rates  that
makes it flexible to be used for different stages of a supernova
explosion. The most important difference of our approach as
compared with the previous ones \cite{Lamb,Lattimer} is that we
consider the whole ensemble of hot primary nuclei, but not only an
average nucleus characterizing the liquid phase. In Fig.~3 we
present the SMM predictions for charge-to-mass ratios (Z/A) and
mass distributions of hot nuclei for typical supernova conditions.
We have performed calculations for different temperatures and
densities at fixed lepton (neutrinos+electrons) to baryon ($Y_L$),
or electron to baryon ($Y_e$) fractions. One can see that at low
temperatures ($T$=1 MeV) the mass distributions have usually three
peaks: at A=1 (free nucleons), A=4 ($\alpha$-particles) and a
large A$\sim$100 corresponding to heavy nuclei. This is typical
picture for a gas-liquid coexistence region. The distribution of
heavy nuclei in this case can be well approximated by a Gaussian
distribution. However, at higher temperatures ($T$=3 MeV) the mass
distributions can be very broad (see dashed line in the bottom
left panel), and they become closer to a power law (for A$>$4). As
seen from the top left panel, the Z/A ratios vary from about 0.45
at low density to about 0.25 at higher densities.

We stress that a great variety of neutron-rich, and even exotic
(large mass---small charge) nuclei can be formed during the explosion.
Our analysis \cite{Botvina04} shows that decreasing the symmetry energy has a
strong influence on the nuclear composition, and favors formation of
neutron-rich nuclei. Therefore, properties of hot neutron-rich nuclei
in supernova environments should be re-considered in accordance with
most recent laboratory experiments, in particular, regarding the
symmetry energy and level densities.
\begin{figure*}[htb!]
\vspace{-1cm}
\includegraphics[width=16cm]{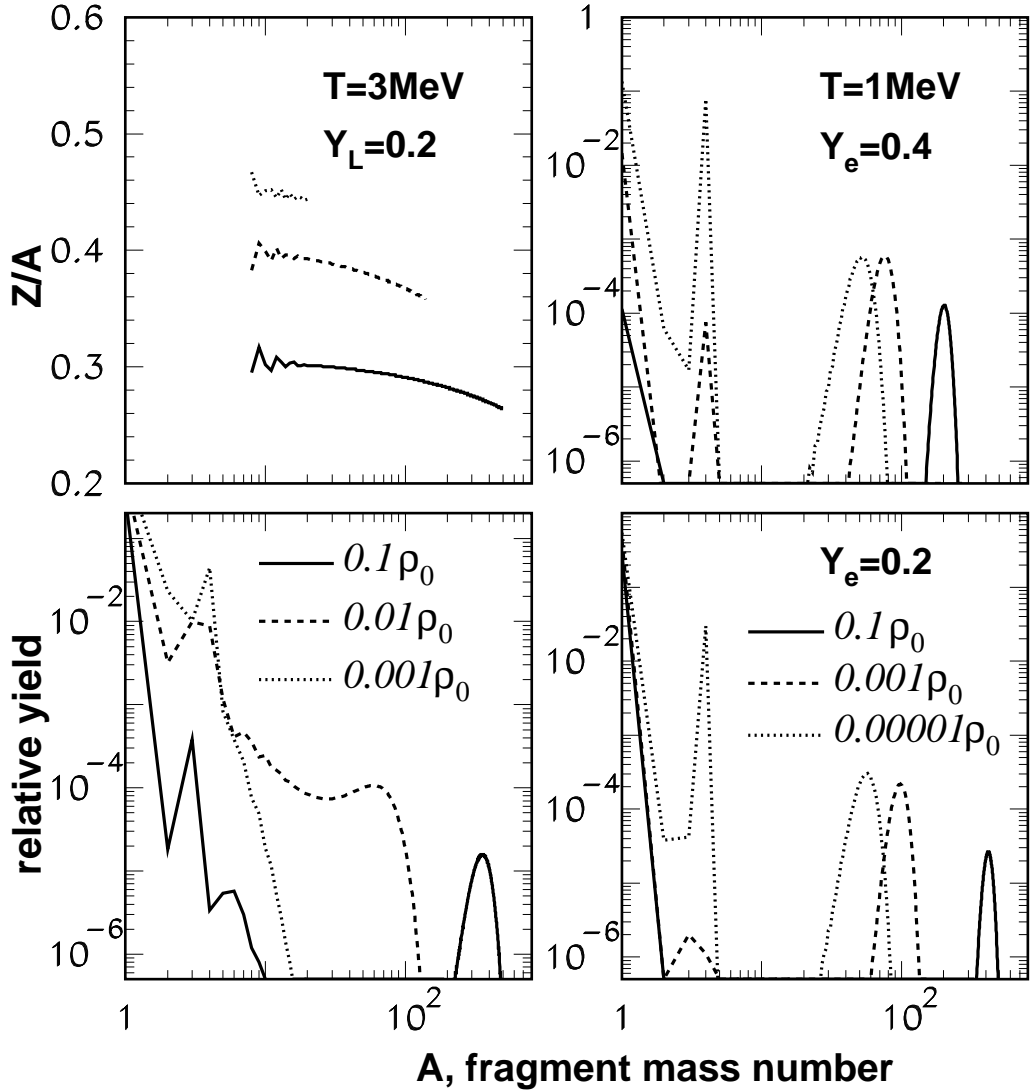}
\caption{\small{ Mean charge-to-mass ratios (left top panel), and
mass distributions of hot primary fragments (other panels)
calculated with the SMM generalized for supernova conditions. Left
panels are calculations for temperature $T=3$ MeV and fixed lepton
(electrons+neutrinos) fraction $Y_L=$0.2 per nucleon. Right panels
are calculations for temperature $T=1$ MeV and fixed electron
fraction $Y_e$ assuming that neutrinos escape. Lines correspond to
baryon densities (in units of the normal nuclear density
$\rho_0$=0.15 fm$^{-3}$) shown in the figure. }}
\end{figure*}


Besides of a more realistic EOS there are other important aspects
of supernova dynamics which are sensitive to the ensemble of hot nuclei.
In particular, the neutrino opacity \cite{Horowitz} and the electron capture rate
\cite{Langanke03} are very sensitive to the nuclear structure
effects. Our approach can be easily generalized to incorporate
the nuclear shell effects which should play a role at temperatures below
1 MeV, and disappear at high excitation energies. It is very
promising to investigate consequences of new shells at very large neutron
excess and possible production of heavy, or even superheavy, nuclei under
supernova conditions.

Also, the nucleosynthesis in supernova environments may proceed
differently if the ensemble of hot nuclei will provide new 'seeds' for
the r-process. Previous r-process calculations were based, as a rule,
on a limited number of stable 'seed' nuclei \cite{Qian}.
On the other hand, as one can see in Fig.~3, a broad variety of primary hot nuclei
should be produced at the nuclear equilibrium stage during the explosion.
After ejection from the hot and dense environment, they will undergo de-excitation.
As known from nuclear experiments, at initial temperature of 1--3 MeV this
is achieved by evaporating only a few nucleons. We expect that in the course
of evaporation certain isotopes will be enhanced by the shell effects.
At later stages these new cold seed nuclei will be further processed in the
standard s- and r-processes. We believe that this two-stage mechanism of
nucleosynthesis may explain some puzzles in the observed nuclear
abundances.

In conclusion, in this paper we have pointed out a similarity of
nuclear matter states reached in heavy-ion reactions and in
supernova explosions. The information on properties of hot
fragments and their production mechanisms can be used for
construction of a reliable equation of state of stellar matter. It
is important that the improved EOS can also be applied for other
astrophysical processes, where the appropriate densities and
temperatures of the matter occur, for example, in the
neutrino-driven protoneutron star wind during the first seconds
after explosions \cite{Thompson}. We hope that in the future our
improved EOS will be implemented into modern hydrodynamical
simulations of supernova dynamics \cite{Janka}. In this way, it
will be possible to perform more realistic calculations of
supernova explosions and synthesis of heavy elements.

The authors thank K.-H. Langanke, L. Sobotka and W. Trautmann for fruitful
discussions. This work was supported in part by the GSI (Germany) and MIS
grant NSH-1885.2003.2 (Russia).



\begin{thebibliography}{99}

\bibitem{Bethe} H.A. Bethe, {\em Rev. Mod. Phys.} {\bf 62}, 801 (1990).

\bibitem{Janka} H.-T. Janka, R. Buras, K. Kifonidis, M. Rampp, and T. Plewa,
Review in "Core Collapse of Massive Stars", Fryer, C.L. (ed.) ,
{\em astro-ph/0212314} (2001); H.-T. Janka and E. Mueller,
{\em Astron. Astrophys.} {\bf 306}, 167 (1996).

\bibitem{Thielemann} M. Liebendorfer et al., {\em Nucl. Phys.} {\bf A719},
144c (2003)

\bibitem{Ring} F.K. Sutaria, A. Ray, J.A. Sheikh, and P. Ring,
Astron. Astrophys. {\bf 349}, 135 (1999).

\bibitem{Hix} W.R. Hix et al., {\em Phys. Rev. Lett.} {\bf 91} 201102 (2003).

\bibitem{Langanke} K. Langanke and G. Martinez-Pinedo, {\em Nucl. Phys.}
{\bf A673}, 481 (2000).

\bibitem{Horowitz} C.J. Horowitz, {\em Phys. Rev.} {\bf D55}, 4577 (1997).

\bibitem{Lamb} J.M. Lattimer, C.J. Pethick, D.G. Ravenhall and D.Q. Lamb,
{\em Nucl. Phys.} {\bf A432}, 646 (1985).

\bibitem{Lattimer} J.M. Lattimer and F.D. Swesty,
{\em Nucl. Phys.} {\bf A535}, 331 (1991).

\bibitem{Botvina04} A.S. Botvina and I.N. Mishustin, {\em Phys. Lett.}
{\bf B584}, 233 (2004).

\bibitem{Japan} C. Ishizuka, A. Ohnishi, K. Sumiyoshi, {\em Nucl.Phys.}
{\bf A723}, 517 (2003).

\bibitem{Langanke03} K. Langanke et al., {\em Phys. Rev. Lett.} {\bf 90}
241102 (2003).

\bibitem{Randrup} S.E. Koonin and J. Randrup, Nucl. Phys. {\bf A471},
355c (1987).

\bibitem{Gross} D.H.E. Gross, Rep. Progr. Phys. {\bf 53}, 605 (1990).

\bibitem{SMM} J.P. Bondorf, A.S. Botvina, A.S. Iljinov, I.N. Mishustin and
K. Sneppen, {\em Phys. Rep.} {\bf 257}, 133 (1995).

\bibitem{ALADIN} A.S. Botvina et al., {\em Nucl. Phys.} {\bf A584},
737 (1995).

\bibitem{EOS} R.P. Scharenberg et al., {\em Phys. Rev.} {\bf C64},
054602 (2001).

\bibitem{ISIS} L. Pienkowski et al., {\em Phys. Rev.} {\bf C65},
064606 (2002).

\bibitem{MSU} M. D'Agostino, et al., {\em Phys. Lett.} {\bf B371}, 175 (1996).

\bibitem{INDRA} N. Bellaize, et al., {\em Nucl. Phys.} {\bf A709}, 367 (2002).

\bibitem{FAZA} S.P. Avdeyev et al., {\em Nucl. Phys.} {\bf A709}, 392 (2002).

\bibitem{Pochodzala} J. Pochodzala and ALADIN collaboration, {\em Phys. Rev.
Lett.} {\bf 75}, 1040 (1995).

\bibitem{Dagostino2} M. D'Agostino, A.S. Botvina, M. Bruno, A. Bonasera,
J.P. Bondorf, I.N. Mishustin et al., {\em Nucl. Phys.} {\bf A650}, 329 (1999).

\bibitem{Bugaev} K.A. Bugaev et al., {\em Phys. Rev.} {\bf C62},
0444320 (2000).


\bibitem{SJNP85} A.S. Botvina, A.S. Iljinov, and I.N. Mishustin, {\em Sov. J. Nucl. Phys.}
{\bf 42}, 712 (1985).

\bibitem{Sobotka}  L.G. Sobotka, R.J. Charity, J. Toke, W. U. Schroder,
Phys. Rev. Lett. {\bf 93}, 132702 (2004).

\bibitem{Botvina01} A.S. Botvina, I.N. Mishustin, {\em Phys. Rev.} {\bf C63},
061601 (2001).

\bibitem{Shetty05} D.V. Shetty et al., {\em Phys. Rev.} {\bf C71},
024602 (2005).

\bibitem{FRS} K.H. Schmidt et al., {\em Nucl. Phys.} {\bf A710}, 157 (2002).

\bibitem{Botvina02} A.S. Botvina, O.V. Lozhkin and W. Trautmann,
             {\em Phys. Rev.} {\bf C65}, 044610 (2002).

\bibitem{LeFevre} A. Le Fevre et al., {\em Phys. Rev. Lett.} {\bf 94}
162701 (2005).

\bibitem{nihal} N. Buyukcizmeci et al., nucl-th/0506017 (2005), to
be published in EPJA.

\bibitem{Qian} Yong-Zhong Qian, Prog. Part. Nucl. Phys. {\bf 50}, 153 (2003).


\bibitem{Thompson} T.A. Thompson, A. Burrows and B.S. Meyer,
{\em Astrophys. J.} {\bf 562}, 887 (2001).

\end{thebibliography}
\end{document}